# Power-Spectrum Analysis of Reconstructed DAMA Data

P.A. Sturrock[a,*], E. Fischbach[b], J.T.Gruenwald[b], D. Javorsek II[c], J.H. Jenkins[b,d], R.F. Lang[b] , R.H. Lee[e], J. Nistor[b] , J. Scargle[f]

[a] Center for Space Science and Astrophysics, Stanford University, Stanford, CA 94305, USA
[b] Department of Physics, Purdue University, West Lafayette, IN 47907, USA
[c] Air Force Test Center, Edwards Air Force Base, CA 93524, USA
[d] Department of Nuclear Engineering, Texas A&M University, College Station, TX 77843, USA
[e] Department of Physics, United States Air Force Academy, Colorado Springs, CO 80920, USA
[f] NASA/Ames Research Center, MS 245-3, Moffett Field, CA 94035, USA
*Corresponding author. Tel +1 6507231438; fax +1 6507234840.
Email address: sturrock@stanford.edu

## Abstract

Claims by the DAMA (DArk MAtter) collaboration to have detected an annually varying signal consistent with models of dark matter appear to be at variance with results from other dark-matter searches. To further understand the DAMA results, we have carried out an independent analysis of DAMA data reconstructed from published figures. In addition to reexamining the Lomb-Scargle and chi-square analyses previously carried out by the DAMA collaboration, we carry out two new likelihood analyses and a new chi-square analysis, focusing attention on the treatment of experimental errors and binning. We confirm the existence of an annual oscillation, with a maximum in early June, but at a lower significance level than previously reported.

## 1 . Introduction

There is great interest in the possibility of detecting dark matter, the existence of which may be inferred from astrophysical data. As a result of the Earth's orbit around the Sun, the incidence of dark-matter particles on Earth is expected to have its maximum value at approximately June 2, when the Earth's orbital velocity is added to that of the Sun with respect to the Galaxy. To search for such an effect, the DAMA (DArk MAtter) collaboration has operated a sequence of two experiments, referred to as DAMA/NaI and DAMA/LIBRA (Large sodium Iodide Bulk for RAre processes). These record signals from highly radiopure sodium-iodide scintillators (100kg of Na(Tl) for DAMA/NaI and 250 kg for DAMA/LIBRA). The DAMA experiments, located at the Gran Sasso National Laboratory, have now been running for over 13 annual cycles with a cumulative exposure of over 1.17 ton-yr. Data published by the DAMA collaboration present evidence (at a claimed confidence level of 8.9 σ) for an annual modulation with a maximum near to June 2. [1-5] Since the DAMA claim that their annual signal originates from WIMP-induced nuclear recoils is at variance with the results from other dark matter searches [6-10], we have undertaken an independent analysis of the DAMA data which we have extracted from their various publications. [1-5]

The DAMA articles contain some ambiguities (as in Table 3 of Ref. [3]) and uncertainties (as, for instance, the exact procedure they used for their power-spectrum analysis). We investigate these issues by our independent analysis of the DAMA data. Since the collaboration has not provided their data in tabular form, it has been necessary to extract their data from eps files of displays in their publications. In Section 2, we present tables of the reconstructed data for the DAMA/NaI and DAMA/LIBRA experiments. We then address the following questions: (a) What is the result of power-spectrum analysis of the DAMA data? and (b) What is the result of chi-square analysis of the DAMA data?

Concerning point (a): we note that the DAMA collaboration has presented the results of power-spectrum analyses for DAMA/NaI data [1], and also for DAMA/LIBRA data and combined DAMA/NaI and DAMA/LIBRA data [3]. These analyses are said to use the Lomb-Scargle procedure but (see, for instance, the caption of Figure 2 of Ref. [3]) taking account of experimental errors and of the aggregation of measurements into bins. The problem is that the basic Lomb-Scargle procedure [11-13] that is referenced in the DAMA article is not designed to take account of either experimental errors or binning. We note as an aside that Scargle has shown how the Lomb-Scargle procedure can be extended to take account of experimental errors (but not bin durations) [14], but this extended procedure is not referenced by, and appears not to have been used by, the DAMA collaboration.

In Section 3, in an attempt to clarify this issue, we carry out analyses of the reconstructed datasets using the basic Lomb-Scargle procedure. In Section 4, we then analyze the datasets using a likelihood power-spectrum procedure that is designed to take account of both experimental errors and bin durations [14].

Concerning point (b): in Section 5 we attempt to understand the discrepancy between the significance level of the annual modulation as inferred from power-spectrum analysis and that proposed by the DAMA collaboration as inferred from a chi-square analysis.

We discuss our results in Section 6.

## 2. DAMA Data

We have reconstructed the experimental results for the DAMA/NaI and DAMA/LIBRA experiments from data published in references [2,3]. Tables 1 and 2 present the 2 – 6 keV data for the DAMA/NaI and DAMA/LIBRA experiments, respectively. For each table, entries in column 2 comprise the dates used by the DAMA collaboration. However, it has been convenient to convert these dates into a format that is very close to calendar dates, and is better suited for time-series analysis.

In our studies of solar neutrinos, we have found it convenient to introduce the term "neutrino days" (column 3), comprising dates counted in days with January 1, 1970, as day 1. We also convert such measurements into "neutrino years" (column 4) as follows:

$$t(Neutrino\,Years) = 1970 + t(Neutrino\,Days)/365.2564 \ . \qquad (1)$$

We have adopted this representation as our standard for the analysis of neutrino and similar data since it has the merit of being a uniformly running measure (it avoids the leap-year problem) that differs only very slightly from the actual calendar date.

We understand that the "residuals," listed in column 5, are to be understood as measurements—in units of counts per day, per kilogram, and per keV—from which the mean value has been subtracted for each annual cycle. Since we are dealing with two distinct experiments, it would have been helpful to be able to normalize each dataset separately with respect to the mean value, but unfortunately the mean values have not been provided.

## 3. Lomb-Scargle Analysis

We now attempt to reproduce the power spectra computed by the DAMA collaboration. The caption of their Figure 2 of ref. [3] reads *Power spectrum of the measured single-hit residuals ...calculated according to Refs. [41,42], including also the treatment of the experimental errors and of the time binning.* The DAMA refs. [41,42] refer to the Lomb-Scargle procedure [11-13]. However, the Lomb-Scargle procedure (described in those articles) does *not* take account of *either* the experimental errors *or* the time binning (i.e. the start and stop times of each measurement), and there is no indication in the DAMA articles that the DAMA collaboration have made any changes in the Lomb-Scargle procedure to take account of these complications.

To investigate this issue, we have applied the Lomb-Scargle procedure to our reconstructed DAMA dataset. Rather than use the equations derived in Refs. [11-13], it is convenient to use a likelihood procedure [15] which, when one assigns a single time to each measurement and adopts the standard deviation of the measurements as the error estimate, is identical to the Lomb-Scargle procedure.

For comparison with Figure 2 of the DAMA article [3], we first apply the Lomb-Scargle procedure to the DAMA/LIBRA data. This yields the power spectrum shown in Figure 1. We follow the DAMA collaboration in computing the power up to the Nyquist frequency, which is approximately 3 year$^{-1}$. In our analysis, the peak power at 1 year$^{-1}$ has the value S = 13.3. The value we find in the left-hand panel of Figure 2 of the DAMA article [3] is 14.2.

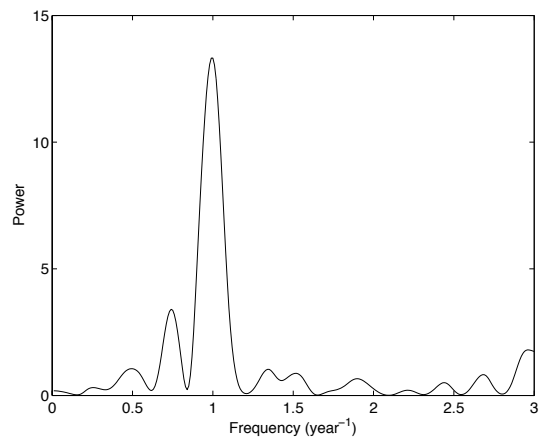

Figure 1. Power spectrum analysis of DAMA/LIBRA data, using the Lomb-Scargle procedure.

For completeness, we have also applied the Lomb-Scargle procedure to the DAMA/NaI data, with the result shown in Figure 3. The peak at 1 year$^{-1}$ now has a power of only 9.1. The corresponding DAMA figure is Figure 4 of Ref. [2], which has a peak value of 10.0.

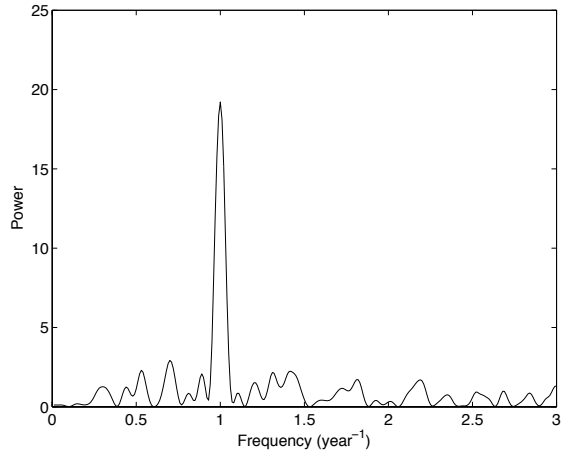

Figure 2. Power spectrum analysis of the combined DAMA/NaI and DAMA/LIBRA data, using the Lomb-Scargle procedure.

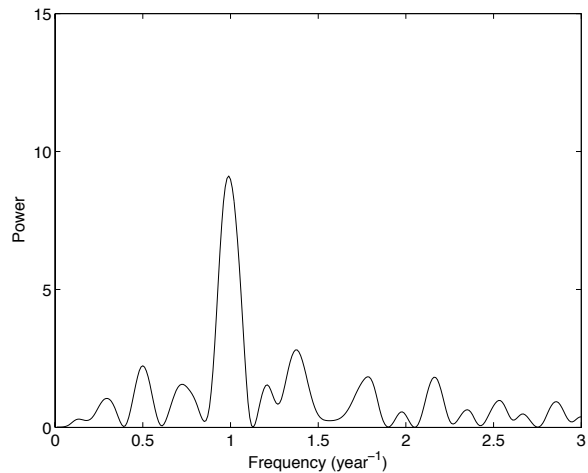

Figure 3. Power spectrum analysis of the DAMA/NaI data, using the Lomb-Scargle procedure.

These comparisons tend to support the hypothesis that the DAMA analysis was a simple Lomb-Scargle calculation that does not take account of either the experimental errors or the time binning. (The analysis in Section 4 will further support this hypothesis.) Our estimates of the power of the annual modulation are lower than the DAMA estimates by a factor of order 10%. We find that random errors of order 1% in our reconstructed DAMA data can explain this discrepancy. On the other hand, random errors of much more than 1% would be incompatible with the fact that our estimates are less than the DAMA estimates by only about 10%, suggesting that our reconstruction of the DAMA data is relatively accurate.

According to Lomb-Scargle theory, there is a probability of $e^{-S}$ of finding a power S or more at a specified frequency. If we were to assume that the concatenated time-series conforms to the requirements of Lomb-Scargle theory (specifically that the errors conform to a standard normal distribution), we could infer that there is a probability of only $e^{-19.2}$, i.e. 4.6 e-9, of finding such a large peak at the (previously specified) frequency of 1.00 year$^{-1}$. A $P$-value of 4.6 e-9 is equivalent to a 5.8 σ confidence level.

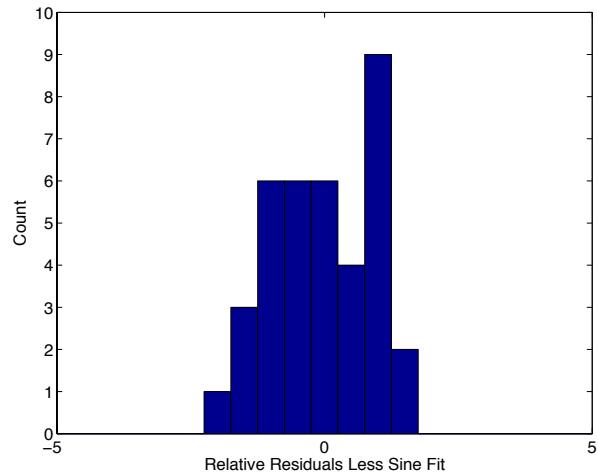

Figure 4. Histogram of residual measurements, less the annual modulation, divided by the residual error estimates, for the DAMA/NaI dataset.

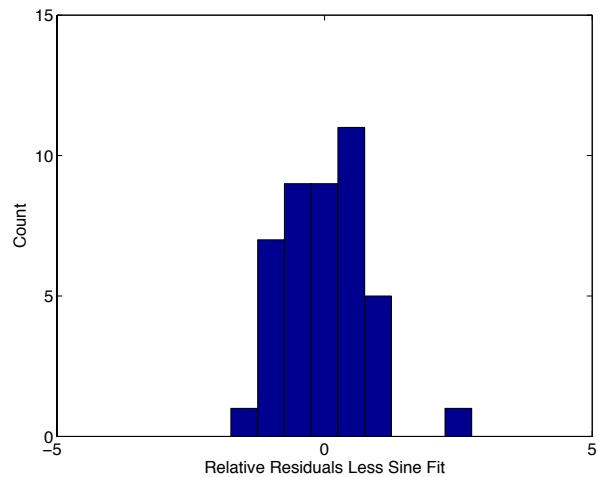

Figure 5. Histogram of residual measurements, less the annual modulation, divided by the residual error estimates, for the DAMA/LIBRA dataset.

In order to determine whether or not the errors are actually distributed normally, we show in Figure 4 a histogram of the ratio of the Residuals less the annual modulation, divided by the Residual Errors, for the DAMA/NaI dataset. We show in Figure 5 the corresponding plot for the DAMA/LIBRA dataset. It would appear that neither is close enough to a normal distribution with standard deviation unity to warrant using significance estimates that are based on those assumptions. Furthermore, the two plots differ significantly in that the standard deviations of the data in the histograms are 0.96 for DAMA/NaI and 0.75 for DAMA/LIBRA. For the valid application of Lomb-Scargle theory, the standard deviations should be unity.

## 4. Likelihood Analysis

The DAMA collaboration proposed (see, for instance, Ref. [3]) the goal of carrying out a power-spectrum analysis that takes account of the experimental errors and of the finite durations of the bins in which counts have been aggregated. The bin durations range from 19.2 days to 98.6 days, with a mean value of 60.0 days. We now carry out such an analysis using a likelihood procedure described elsewhere [15].

The result of this analysis is shown in Figure 6. The peak at 1.00 year$^{-1}$ has a power of 37.1. If interpreted as a valid power-spectrum analysis, we would infer that there is a probability of only 7.7e-17 of finding a peak this big or more by chance, which would convert to a confidence level of 8.2 σ.

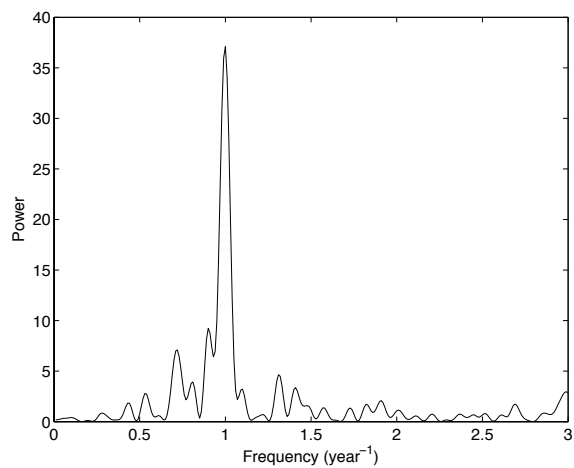

Figure 6. Power spectrum formed from the combined DAMA/NaI and DAMA/LIBRA data, using a likelihood procedure to take account of the experimental errors (accepted at face value) and of the binning. The peak at 1.00 year$^{-1}$ has power $S = 37.10$.

However, this calculation depends critically on the assumption that each measurement may be viewed as a sample drawn from a distribution of hypothetical measurements which has the form of a normal distribution with a standard deviation set by the error estimate. Without knowing precisely how the error estimates are arrived at, we have no way of knowing whether or not this assumption is justified. Furthermore, we noted in the previous section that there is cause for concern about whether error estimates may be accepted at face value.

For this reason, it is advisable to obtain a significance estimate that does not depend on that assumption. One suitable procedure is the shuffle test [16]. We adopt one list that comprises the start time and end time of each bin, and another list that comprises the Residual and Residual Error of each bin. We then randomly associate time-pairs from the former list with the measurement pairs from the latter list, shuffling each dataset separately. We have carried out this procedure 100,000 times and arrived at the distribution of power measurements (at the fixed frequency 1.00 year$^{-1}$) shown in logarithmic form in Figure 7.

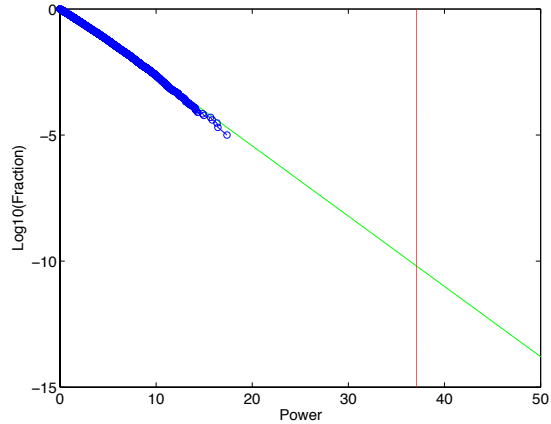

Figure 7. Logarithmic display of the power at 1.00 year$^{-1}$ as a result of 100,000 shuffle simulations, each taking account of the experimental errors estimates and the binning. The fraction with a power of 37.10 or more is $10^{-10.18}$. This corresponds to an equivalent power of 23.44, which converts to 6.46 σ.

On projecting the curve to S = 37.10, we estimate that the probability of obtaining a power of 37.10 or more by chance is 6.6e-11. This corresponds to an equivalent power of 23.4, which leads to a significance estimate of 6.5 σ, a somewhat stronger result than we obtained from the Lomb-Scargle analysis (5.8 σ), but more conservative than the estimate of 8.9 σ proposed by the DAMA Collaboration [4].

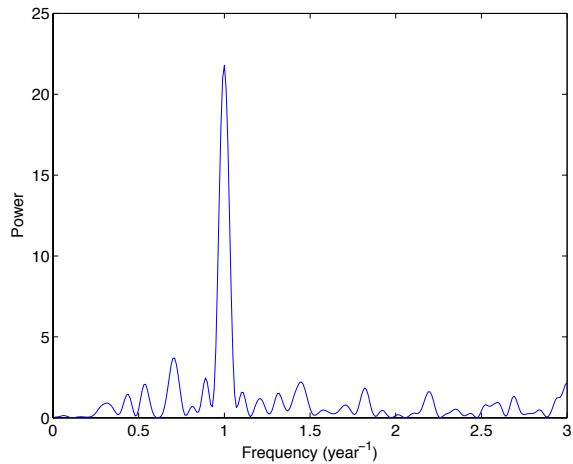

Figure 8. Power spectrum formed by a likelihood procedure, which is equivalent to the Lomb-Scargle procedure, after applying the rono operation to the DAMA/NaI and DAMA/LIBRA datasets separately, and then combining them. The peak at 1.00 year$^{-1}$ has power S = 21.80.

We have recently introduced another procedure for likelihood power-spectrum analyses, which we term the *rono* (rank-order normalization operation) procedure [17]. According to this procedure, we map measurements onto a normal distribution with standard deviation unity, retaining the order of measurements. The advantage of this procedure, in the current context, is that we may apply this operation to the two datasets independently, so that they then have exactly the same statistical properties, and then combine them. Furthermore, these properties are exactly those that are appropriate for the application of the Lomb-Scargle procedure. The resulting spectrum is shown in Figure 8. The power of the annual modulation is found to be 21.8, corresponding to a P-Value of 3.4e-10 and a confidence level of 6.1 σ.

## 5. Chi-Square Analysis

The DAMA collaboration has presented the results of a chi-square test, but we find the last line of Table 3 of Ref. [3] to be puzzling. This line indicates that a chi-square value of 64.7, for 79 degrees of freedom, leads to 8.8 σ as the confidence level for the presence of an annual modulation. However, we find that this combination leads to the inference that there is a probability of 0.88 that these figures are consistent with an annual modulation (or a probability of 0.12 that the figures are inconsistent with that assumption). This leads us to suspect that there is a typographical error in the table. If the number of degrees of freedom is taken to be 2, as is appropriate if one allows for uncertainty in both amplitude and phase (the mean value of the offset is zero by definition), we find that the probability of no annual modulation (fixing the frequency) is 8.9e-15, which corresponds to a 7.7 σ result.

On the other hand, the caption to the DAMA figure indicates that the phase has been set to $t_0$ = 152.5 days. If we remove this degree of freedom, the probability of no annual modulation becomes 8.9e-16, which corresponds to a 7.9 σ result. (However, it is difficult to reconcile the fact that—according to the caption of Table 3 of Ref. [3]—the phase has been fixed, with the statement in the abstract that the collaboration has obtained a "model independent" result.)

Neither of the above significance estimates agrees with the one given in Table 3 of Ref [3], which is 8.8 σ C.L. Moreover, the above estimates of 7.7 σ and 7.9 σ differ significantly from the estimates we have found in Sections 3 and 4. We therefore investigate the chi-square estimate further.

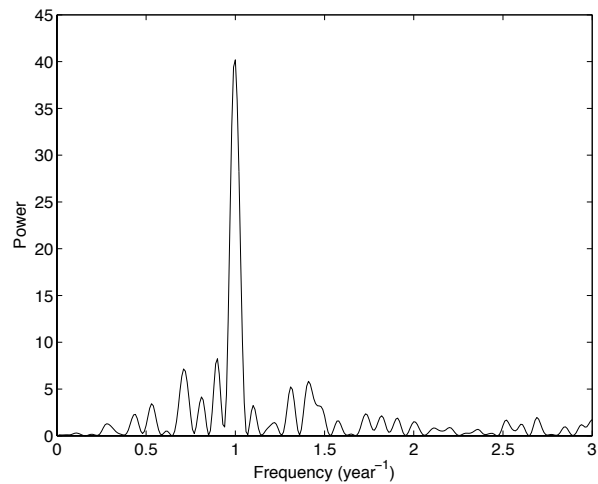

Figure 9. Power spectrum formed from the combined DAMA/NaI and DAMA/LIBRA data, using a likelihood procedure that takes account of the errors but assigns a single time to each bin. The peak at 1.00 year[-1] has power S = 40.19.

If we adopt 2 degrees of freedom (as is appropriate if we wish to allow for uncertainty in both amplitude and phase), the chi-square statistic is simply twice the power at the given frequency, as it would be computed taking account of the experimental error, but not taking account of binning. We therefore carry out such a likelihood power spectrum analysis. The result is shown in Figure 9. The power at 1.00 year$^{-1}$ is 40.2, not very different from the value 37.1 derived in Section 4 from a likelihood analysis that takes account of the binning and adopts the experimental errors at face value. S = 40.2 corresponds to $P = 3.5e - 18$, $\chi^2 = 80.4$, and 8.6 σ C.L. This is not far from the value 8.8 σ C.L. listed in Table 3 of Ref. [3].

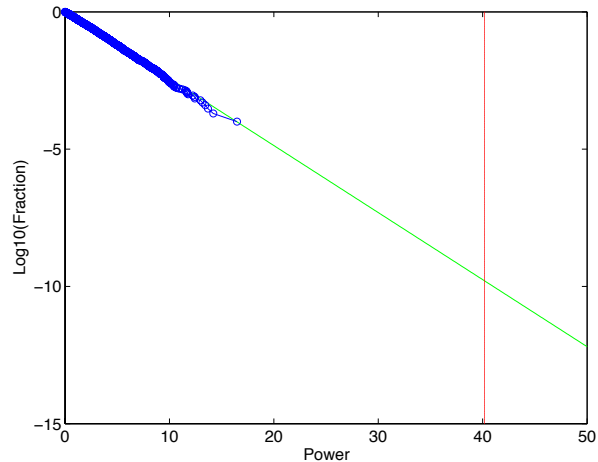

Figure 10. Logarithmic display of the power at 1.00 year$^{-1}$ as a result of 10,000 shuffle simulations, each taking account of the experimental error estimates but not of the binning. The fraction with a power of 40.19 or more is $10^{-9.73}$. This corresponds to an equivalent power of 22.40, which converts to 6.3 σ.

We show in Figure 10, in logarithmic form, the result of 10,000 shuffles of the data, followed by the above power-spectrum analysis. We see that the fraction with a power of 40.19 or more is $10^{-9.73}$. This corresponds to an equivalent power of 22.40, which converts to 6.3 σ, not far from the values found in Sections 3 and 4, but more conservative than the value (8.8 σ) shown in Table 3 of Ref. [3] as a result of a chi-square analysis by the DAMA collaboration.

## 6. Discussion

We have obtained reasonably consistent results from three versions of power-spectrum analysis: a Lomb-Scargle analysis, an analysis that uses the rank-order-normalization procedure, and one that uses the shuffle test. These procedures all lead to evidence for an annual modulation with confidence levels ranging from 5.8 σ to 6.6 σ. We also argue that a chi-square analysis yields a confidence level in this range. This is strong evidence for an annual oscillation, but not as strong as confidence levels estimated by the DAMA collaboration, which range as high as 8.9 σ (in the abstract of ref. [3]). The DAMA estimates rest critically on their experimental error estimates that (as pointed out in Section 3) do not have a standard normal form.

It is unfortunate that the DAMA analyses have focused exclusively on the possibility of an annual oscillation (potentially related to dark matter). The fact that measurements have been aggregated in bins of about 60-day duration has the consequence that one cannot examine the power spectrum for frequencies much above 3 year$^{-1}$. Furthermore, the fact that the DAMA collaboration calculates the residuals by subtracting the mean on a yearly basis [18] has the consequence that one cannot search for periodicities with frequencies less than 1 year$^{-1}$. Since individual events can presumably be registered with times of less than a second, there is in principle no problem in scanning an extensive frequency range, such as 0 to 100 year$^{-1}$ or more. This would make it possible to easily search for oscillations with frequencies less than 1 year$^{-1}$, for harmonics of the annual oscillation, and for possible solar rotational signals (in the band 9 – 14 year$^{-1}$ [19,20]). Examining unbinned data would also make it possible to search for a possible diurnal variation. (The SNO (Sudbury Neutrino Observatory) collaboration, for instance, took advantage of the accurate timing of their neutrino-capture events to carry out such a wide-band power-spectrum analysis [21].) By focusing on a very small frequency band, the collaboration may be missing important information encrypted in their data that may be relevant to dark matter, and may also be relevant to alternative interpretations of DAMA experimental results, such as a possible role of muons (suggested by Ralston [22], Nygren [23] and Blum [24]) or of K40 decays (suggested by Pradler et al. [25,26] but criticized by Bernabei et al. [27]), or to a possible solar influence.

It is unfortunate that the DAMA collaboration has packaged their data in such a way that it is possible to examine only a very small range for frequencies. It is to be hoped that the collaboration will soon make their raw (unpackaged) data available to the scientific community.

Table 1. Reconstructed data for the DAMA/NaI experiment.

| Line Number | Date (DAMA Notation) | Date (Neutrino Days) | Date (Neutrino Years) | Residual | Day Error | Residual Error |
|---|---|---|---|---|---|---|
| 1 | 355 | 9486 | 1995.971 | -0.012 | 44.1 | 0.012 |
| 2 | 529.7 | 9660.7 | 1996.449 | 0.036 | 15.7 | 0.02 |
| 3 | 739.3 | 9870.3 | 1997.023 | -0.026 | 49.3 | 0.011 |
| 4 | 809.2 | 9940.2 | 1997.214 | 0.002 | 19.7 | 0.015 |
| 5 | 849.3 | 9980.3 | 1997.324 | 0.033 | 19.7 | 0.013 |
| 6 | 889.5 | 10020.5 | 1997.434 | 0.009 | 19.7 | 0.013 |
| 7 | 924.5 | 10055.5 | 1997.53 | -0.007 | 14.8 | 0.019 |
| 8 | 966.4 | 10097.4 | 1997.645 | -0.026 | 27.1 | 0.013 |
| 9 | 1029.3 | 10160.3 | 1997.817 | -0.038 | 34.9 | 0.013 |
| 10 | 1109.6 | 10240.6 | 1998.037 | 0 | 45.4 | 0.01 |
| 11 | 1169 | 10300 | 1998.199 | 0.015 | 14.8 | 0.019 |
| 12 | 1209.2 | 10340.2 | 1998.309 | 0.033 | 24.9 | 0.016 |
| 13 | 1265.1 | 10396.1 | 1998.462 | 0.016 | 29.7 | 0.013 |
| 14 | 1327.9 | 10458.9 | 1998.635 | 0.017 | 32.3 | 0.012 |
| 15 | 1399.6 | 10530.6 | 1998.831 | 0 | 29.7 | 0.016 |
| 16 | 1474.7 | 10605.7 | 1999.036 | -0.019 | 45.4 | 0.011 |
| 17 | 1534.1 | 10665.1 | 1999.199 | -0.004 | 14.8 | 0.018 |
| 18 | 1579.5 | 10710.5 | 1999.323 | -0.005 | 29.7 | 0.014 |
| 19 | 1644.1 | 10775.1 | 1999.5 | 0.033 | 34.9 | 0.013 |
| 20 | 1693 | 10824 | 1999.634 | 0.017 | 9.6 | 0.028 |
| 21 | 1734.9 | 10865.9 | 1999.749 | -0.009 | 28.4 | 0.019 |
| 22 | 1789.1 | 10920.1 | 1999.897 | -0.035 | 24.9 | 0.016 |
| 23 | 1859 | 10990 | 2000.088 | 0.003 | 44.1 | 0.01 |
| 24 | 1944.5 | 11075.5 | 2000.323 | 0.016 | 40.2 | 0.011 |
| 25 | 2014.4 | 11145.4 | 2000.514 | -0.001 | 29.7 | 0.015 |
| 26 | 2145.4 | 11276.4 | 2000.873 | -0.005 | 14.8 | 0.019 |
| 27 | 2175.1 | 11306.1 | 2000.954 | -0.015 | 15.7 | 0.015 |
| 28 | 2224 | 11355 | 2001.088 | -0.004 | 34.9 | 0.011 |
| 29 | 2285.2 | 11416.2 | 2001.255 | 0 | 25.8 | 0.012 |
| 30 | 2328.8 | 11459.8 | 2001.375 | 0.014 | 20.1 | 0.014 |
| 31 | 2374.2 | 11505.2 | 2001.499 | 0.009 | 24.9 | 0.013 |
| 32 | 2431.9 | 11562.9 | 2001.657 | -0.015 | 31.9 | 0.013 |
| 33 | 2494.8 | 11625.8 | 2001.829 | -0.017 | 29.7 | 0.011 |
| 34 | 2564.6 | 11695.6 | 2002.02 | -0.006 | 40.2 | 0.01 |
| 35 | 2645 | 11776 | 2002.24 | 0.004 | 40.2 | 0.011 |
| 36 | 2699.1 | 11830.1 | 2002.389 | 0.037 | 14.8 | 0.017 |
| 37 | 2734.1 | 11865.1 | 2002.484 | 0.029 | 19.7 | 0.017 |

Table 2. Reconstructed data for the DAMA/LIBRA experiment.

| Line Number | Date (DAMA Notation) | Date (Neutrino Days) | Date (Neutrino Years) | Residual | Day Error | Residual Error |
|---|---|---|---|---|---|---|
| 1 | 3195.2 | 12326.2 | 2003.747 | -0.006 | 28.4 | 0.006 |
| 2 | 3250 | 12381 | 2003.897 | -0.011 | 24.8 | 0.008 |
| 3 | 3300 | 12431 | 2004.034 | -0.008 | 24.8 | 0.007 |
| 4 | 3350.1 | 12481.1 | 2004.171 | 0.008 | 24.8 | 0.007 |
| 5 | 3395.6 | 12526.6 | 2004.295 | 0.01 | 20 | 0.008 |
| 6 | 3430.2 | 12561.2 | 2004.39 | 0.008 | 14.8 | 0.009 |
| 7 | 3460.2 | 12591.2 | 2004.472 | 0 | 14.6 | 0.009 |
| 8 | 3523 | 12654 | 2004.644 | -0.006 | 48 | 0.006 |
| 9 | 3584.8 | 12715.8 | 2004.814 | -0.016 | 14.8 | 0.01 |
| 10 | 3614.9 | 12745.9 | 2004.896 | -0.011 | 14.8 | 0.01 |
| 11 | 3654.9 | 12785.9 | 2005.005 | -0.005 | 24.8 | 0.007 |
| 12 | 3710.4 | 12841.4 | 2005.157 | 0.001 | 30 | 0.007 |
| 13 | 3795 | 12926 | 2005.389 | 0.011 | 24.8 | 0.009 |
| 14 | 3835.1 | 12966.1 | 2005.499 | 0.017 | 14.8 | 0.01 |
| 15 | 3865.1 | 12996.1 | 2005.581 | 0.005 | 14.1 | 0.011 |
| 16 | 3897.9 | 13028.9 | 2005.671 | 0.004 | 18.9 | 0.009 |
| 17 | 3940.6 | 13071.6 | 2005.788 | 0.013 | 24.6 | 0.008 |
| 18 | 3979.8 | 13110.8 | 2005.895 | -0.017 | 15 | 0.009 |
| 19 | 4019.8 | 13150.8 | 2006.004 | -0.01 | 24.8 | 0.006 |
| 20 | 4069.8 | 13200.8 | 2006.141 | -0.002 | 24.8 | 0.007 |
| 21 | 4115.3 | 13246.3 | 2006.266 | 0.004 | 20 | 0.007 |
| 22 | 4159.9 | 13290.9 | 2006.388 | 0.013 | 24.8 | 0.007 |
| 23 | 4210 | 13341 | 2006.525 | 0.002 | 24.8 | 0.007 |
| 24 | 4267.3 | 13398.3 | 2006.682 | -0.007 | 32.1 | 0.008 |
| 25 | 4320.1 | 13451.1 | 2006.826 | -0.005 | 20 | 0.007 |
| 26 | 4360.1 | 13491.1 | 2006.936 | -0.016 | 20 | 0.007 |
| 27 | 4405.6 | 13536.6 | 2007.061 | -0.002 | 25 | 0.006 |
| 28 | 4460.2 | 13591.2 | 2007.21 | 0.006 | 30 | 0.006 |
| 29 | 4514.8 | 13645.8 | 2007.36 | 0.014 | 24.8 | 0.007 |
| 30 | 4560.3 | 13691.3 | 2007.484 | 0.003 | 20 | 0.008 |
| 31 | 4613.1 | 13744.1 | 2007.629 | -0.004 | 32.3 | 0.005 |
| 32 | 4674.9 | 13805.9 | 2007.798 | -0.004 | 30 | 0.005 |
| 33 | 4725 | 13856 | 2007.935 | -0.01 | 20 | 0.007 |
| 34 | 4780.5 | 13911.5 | 2008.087 | 0.001 | 35.3 | 0.005 |
| 35 | 4840.5 | 13971.5 | 2008.251 | 0.008 | 25 | 0.006 |
| 36 | 4890.6 | 14021.6 | 2008.388 | 0.014 | 25 | 0.006 |
| 37 | 4935.2 | 14066.2 | 2008.51 | 0.004 | 20 | 0.007 |
| 38 | 4973.4 | 14104.4 | 2008.615 | -0.007 | 18.4 | 0.009 |
| 39 | 5079.8 | 14210.8 | 2008.907 | -0.011 | 20 | 0.006 |
| 40 | 5135.4 | 14266.4 | 2009.059 | -0.007 | 35.3 | 0.004 |
| 41 | 5205.4 | 14336.4 | 2009.25 | -0.001 | 35.3 | 0.004 |
| 42 | 5275.5 | 14406.5 | 2009.442 | 0.01 | 35.3 | 0.004 |
| 43 | 5333.7 | 14464.7 | 2009.602 | 0.009 | 23.7 | 0.005 |